\documentclass{aastex631}

\newcommand{\fastfrb}{FRB 20190520B}
\shorttitle{PRS associated with \fastfrb}
\shortauthors{Zhang et al.}
\graphicspath{{./}{figures/}}
\begin{document}

\title{Temporal and Spectral Properties of the Persistent Radio Source Associated with \fastfrb~with the VLA}

\author[0000-0002-8086-4049]{Xian Zhang}
\affiliation{Shanghai Astronomical Observatory, Chinese Academy of Sciences, 80 Nandan Road, Shanghai 200030, China}
\affiliation{University of Chinese Academy of Sciences, 19A Yuquanlu, Beijing 100049, China}

\author[0000-0002-3844-9677]{Wenfei Yu}
\affiliation{Shanghai Astronomical Observatory, Chinese Academy of Sciences, 80 Nandan Road, Shanghai 200030, China}

\author{Casey Law}
\affiliation{Cahill Center for Astronomy and Astrophysics, MC 249-17 California Institute of Technology, Pasadena, CA 91125, USA}
\affiliation{Owens Valley Radio Observatory, California Institute of Technology, 100 Leighton Lane, Big Pine, CA, 93513, USA}

\author{Di Li}
\affiliation{National Astronomical Observatories, Chinese Academy of Sciences, Beijing 100012, China}
\affiliation{University of Chinese Academy of Sciences, 19A Yuquanlu, Beijing 100049, China}
\affiliation{Research Center for Intelligent Computing Platforms, Zhejiang Laboratory, Hangzhou 311100, China}

\author{Shami Chatterjee}
\affiliation{Cornell Center for Astrophysics and Planetary Science, and Department of Astronomy, Cornell University, Ithaca, NY, USA.}

\author{Paul Demorest}
\affiliation{National Radio Astronomy Observatory, 1003 Lopezville Rd., Socorro, NM 87801, USA}

\author{Zhen Yan}
\affiliation{Shanghai Astronomical Observatory, Chinese Academy of Sciences, 80 Nandan Road, Shanghai 200030, China}

\author{Chenhui Niu}
\affiliation{National Astronomical Observatories, Chinese Academy of Sciences, Beijing 100012, China}
\affiliation{Institute of Astrophysics, Central China Normal University, Wuhan 430079, China}

\author{Kshitij Aggarwal}
\affiliation{Department of Physics and Astronomy, West Virginia University, Morgantown, WV 26506, USA}
\affiliation{Center for Gravitational Waves and Cosmology, West Virginia University, Morgantown, WV 26506, USA}

\author{Reshma Anna-Thomas}
\affiliation{Department of Physics and Astronomy, West Virginia University, Morgantown, WV 26506, USA}
\affiliation{Center for Gravitational Waves and Cosmology, West Virginia University, Morgantown, WV 26506, USA}

\author{Sarah Burke-Spolaor}
\affiliation{Department of Physics and Astronomy, West Virginia University, Morgantown, WV 26506, USA}
\affiliation{Center for Gravitational Waves and Cosmology, West Virginia University, Morgantown, WV 26506, USA}

\author{Liam Connor}
\affiliation{Cahill Center for Astronomy and Astrophysics, MC 249-17 California Institute of Technology, Pasadena, CA 91125, USA}

\author{Chao-Wei Tsai}
\affiliation{National Astronomical Observatories, Chinese Academy of Sciences, Beijing 100012, China}

\author{Weiwei Zhu}
\affiliation{National Astronomical Observatories, Chinese Academy of Sciences, Beijing 100012, China}

\author[0000-0002-1583-8514]{Gan Luo}
\affiliation{National Astronomical Observatories, Chinese Academy of Sciences, Beijing 100012, China}
\affiliation{School of Astronomy and Space Science, Nanjing University, Nanjing 210093, China}



\begin{abstract}
Among more than 800 known fast radio bursts (FRBs), only two, namely FRB 20121102A and FRB 20190520B, are confirmed to be associated with persistent radio sources (PRSs). Here we report evidence of apparent temporal variability in the PRS associated with the bursting \fastfrb~based on the Karl G. Jansky Very Large Array (VLA) observations taken in 2020 and 2021. Based on the analysis of epoch-to-epoch variability of the PRS at L, S, C, and X band (1--12~GHz), we detected not only overall marginal variability but also a likely radio flux decrease ($\sim$3.2 $\sigma$) between the observations taken in 2020 and 2021 at 3~GHz. Assuming no spectral variation in the PRS during these observations, we found the evidence for an overall broad-band radio flux decrease by about 20\% between the 2020 and the 2021 observations, suggesting that the PRS probably evolves on the yearly time scale. If we attribute the marginal variability at 3~GHz as intrinsic or due to scintillation, the size of potential variable component of the PRS is constrained to be sub-parsec. On the other hand, the size of the PRS can be also constrained to $\gtrsim$~0.22~pc from the time-averaged radio spectrum and the integrated  radio luminosity in the 1--12~GHz band based on equipartition and self-absorption arguments. We discuss potential origins of the PRS and suggest that an accreting compact object origin might be able to explain the PRS's temporal and spectral properties. Confirmation of variability or flux 
decline of the PRS would be critical to our understanding of the PRS and its relation to the bursting source.

\end{abstract}



\section{Introduction} \label{sec:intro}
Fast Radio Bursts (FRBs) are extremely bright radio flashes of $\sim$ millisecond duration and high brightness temperature ($>$ 10$^{32}$K), indicating a coherent emission mechanism from a compact region \citep{2019A&ARv..27....4P}. The exact underlying physical mechanism is still under debate, with likely emission mechanism ranging from magnetospheric to shock emission \citep[e.g.,][]{2020Natur.587...45Z}, produced by injection of relativistic outflow into a strongly magnetized medium. FRBs are believed to be associated with compact objects such as magnetars \citep{2013arXiv1307.4924P, 2020Natur.587...59B}, pulsars \citep{2016ApJ...829...27D}, X-ray Binaries \citep{2017MNRAS_Katz,2021ApJ...917...13S,2021ApJ...922...98D,2022ApJ_Sridhar}. There are at least 34~\citep{2020ApJ...903..152H,gordon2023demographics,law2023deep} localized FRBs of the FRB source sample ($>$~800\footnote{\url{https://www.wis-tns.org}}) as recorded in the Supernova Working Group Transient Name Server (TNS). Precise localization of FRBs provides meaningful insights about them by enabling us to identify any multi-wavelength counterpart(s) and by revealing information about their central engine and surrounding environment \citep{2018Natur.553..182M, 2022Sci...375.1266F}. %Only 10 such repeating FRBs has been localized so far.}. %critical clues to their origins \citep{2013arXiv1307.4924P, 2017ApJ...841...14M,2022arXiv_Chen}. 
Precise localization of the first-known repeater FRB~20121102A \citep{2017Natur.541...58C} allowed the identification of a star-forming dwarf host galaxy at a redshift $z=0.193$ \citep{2017ApJ...834L...7T}, which is spatially coincident with a persistent radio source (hereafter PRS) that is compact \citep[$<$~0.7 pc;][]{2017ApJ...834L...8M}. As of yet, %only 8 repeating FRBs have been localized to host galaxies \citep[][and references therein]{2022ApJ...927...55L}. 
only limited deep radio continuum observations were made for localized FRBs and only two FRB$-$PRS associations have been found in the sample of localized FRBs with a host galaxy identification \citep[][and references therein]{2022ApJ...927...55L}, which are FRB~20121102A and \fastfrb~\citep{2022Nature_Niu}, confirmed to be associated with a compact PRS. 

\fastfrb~was detected with the Five-hundred-meter Aperture Spherical radio Telescope \citep[FAST;][]{2011_NRD} in drift-scan mode as part of the Commensal Radio Astronomy FAST Survey \citep[CRAFTS;][]{Li18,2019RAA_LiDi} at 1.05--1.45~GHz in 2019 and later it was localized by the Karl G. Jansky Very Large Array (VLA) using the prestigious realfast fast transient detection system \citep{2018ApJS_Law} with the 2020 observations~\citep{2022Nature_Niu}. \fastfrb~resembles FRB~20121102A in its active repeating nature, its association with a compact PRS (only the second such association), in its association with a star-forming dwarf host galaxy, in its larger host Dispersion Measure (DM)~\citep{2022Nature_Niu} in comparison to other FRBs and in its substantial rotation measure (RM)~\citep{2023Sci_Anna}. One possibility is that the large host DM and RM arise in a dense nebula surrounding the central engine of~\fastfrb, as expected from models of FRB emission that invoke young magnetars embedded in their wind nebulae \citep{2017ApJ...841...14M,2018ApJ_Margalit}. If so, the compact PRSs associated with \fastfrb~and FRB~20121102A represent a different stage of evolution of the nebula compared to other FRBs, and their properties  %Such compact PRSs 
would be the key to probe the near-source environment and origin of the FRBs in both theory \citep{2017ApJ...839L...3K,2017ApJ...841...14M,2020ApJ...895....7Y,2022ApJ...928L..16Y} and observation \citep{2018Natur.553..182M,2021A&A...655A.102R,2022arXiv_Chen}. %Recently, \cite{2023MNRAS.tmp.2383R} reported that the PRS associated with FRB 20121102A show a reduction in flux density at 1.3 GHz of about 30\% over three years, which possibly originates from variation in the termination shock. This could be the result of interaction between the pulsar wind and a reverse shock from a supernova.
There have been some FRB host galaxies found to harbour a radio source but later confirmed to be associated with star forming activity~\citep[e.g., FRB 20191001A, FRB 20190608B, FRB 20181030A, FRB 20201124A;][]{2020ApJ...901L..20B,2020ApJ...895L..37B,2021ApJ...919L..24B,2022MNRAS.513..982R}. One example is the repeating FRB~20201124A, which was also spatially associated with a radio source, revealed by the upgraded Giant Metrewave Radio Telescope \citep[uGMRT;][]{2021ATel14529....1W} and the VLA \citep{2021ATel14549....1R}. However, optical spectroscopy and radio interferometric measurements demonstrated that the source was spatially extended on scales $\gtrsim$ 50~mas and consistent with the ongoing star formation activity in the host galaxy \citep{2022MNRAS.513..982R,dong2023mapping}. Notably, a MeerKAT, e-MERLIN coordinated search has led to another FRB$-$PRS association from FRB 20190714A \citep{2022MNRAS.515.1365C}, but it's not clear yet if this radio emission could rather be accounted for by ongoing star forming activity. Up to now, no more FRB--PRS associations have been found with other localized FRBs \citep[][and references therein]{2022ApJ...927...55L}. 
%Four bursts were detected during the initial 24-second scan. Monthly follow-up tracking observations between 2020 April and 2020 September detected 75 bursts in 18.5~hrs with a mean pulse Dispersion Measure (DM) of $1204.7\pm 4.0$~pc~cm$^{-3}$, indicating that the FRB has a high burst rate~\citep{2022Nature_Niu}. 
In this paper, we report on the observational study of the VLA observations of the PRS associated with \fastfrb~in 2020 and 2021, and discuss some possible origins of the PRS of \fastfrb~implied from digging into its detailed temporal and spectral measurements.

\section{VLA Observations and Data Reduction} \label{sec:observations}
\subsection{2020 observations}\label{subsec:20_obs}
The summary of the 2020 VLA observations is listed in Table~\ref{tab:vla_observations} and also see \citet{2022Nature_Niu} for more details. The data flagging and calibration are detailed in \citet{2022Nature_Niu} and the measured source properties (e.g., flux densities and beam size) are incorporated in Table~\ref{tab:vla_observations}. Notice that we have added an additional 5\% systematic fluxscale error for the measured flux density to the corresponding statistical errors, which is typical for analyzing VLA observations \citep{2017ApJS..230....7P}.

\begin{figure*}
\centering
\includegraphics[width=0.95\textwidth]{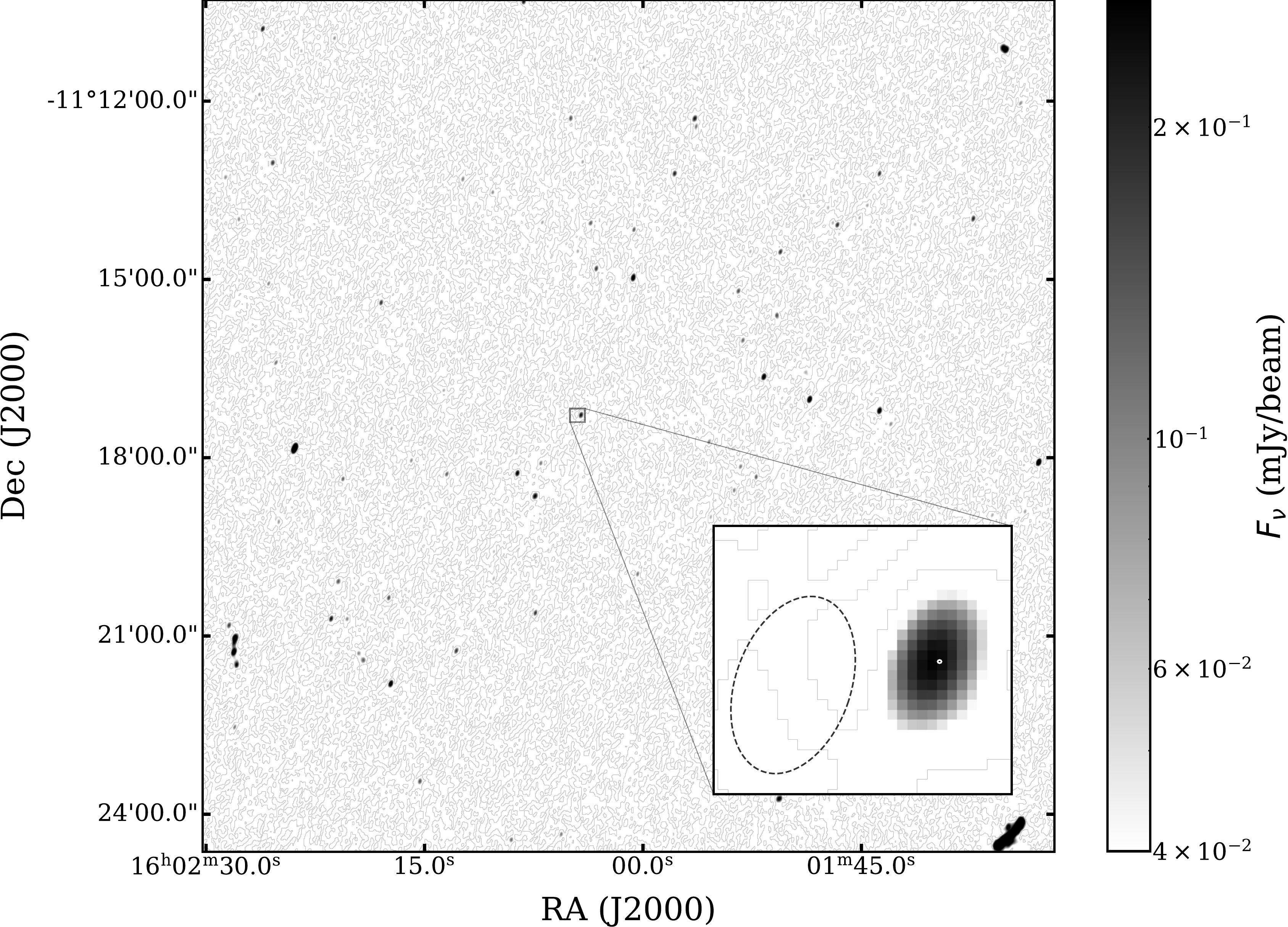}

\caption{Deep image of the PRS associated with \fastfrb~at 1.5~GHz in 2020 VLA observations. Inset shows the image of the PRS, the averaged position of the bursts detected by VLA over-plotted as the white ellipse, and the synthesized beam manually over-plotted to the left of the PRS as an ellipse (the size is $4.7^{\prime \prime}$$\times$$2.9^{\prime \prime}$ and the position angle is -19.2$\arcdeg$). The PRS is spatially and physically associated with \fastfrb~given the chance of coincident association is negligible.}
\label{fig:FRB190520_Lband}
\end{figure*}

Fig.~\ref{fig:FRB190520_Lband} shows the deep image of the PRS at 1.5~GHz in 2020 VLA observations. The PRS has a spectral luminosity of $L_{\rm{1.5~GHz}}=4.7\times10^{29}$\ erg s$^{-1}$\ Hz$^{-1}$ at the luminosity distance of 1218 Mpc revealed from the 1.5~GHz deep image, while the chance of coincident association of the PRS with the FRB position is $\approx3\times{10}^{-6}$~\citep{2022Nature_Niu}, implying the PRS is physically connected with the repeating \fastfrb.

\subsection{2021 observations}\label{subsec:21_obs}
In order to investigate the temporal and spectral properties of the PRS associated with \fastfrb, we proposed and obtained two epochs of additional VLA observations of the source for two epochs on 2021 October 1 and November 7 (Project ID: 21B-127; PI: Prof. Wenfei Yu), while VLA/realfast was not equipped during the observations. Each observation consists of five radio bands which were taken one-by-one: P (0.35~GHz), L (1.5~GHz), S (3~GHz), C (5.5~GHz), and X (10~GHz) bands. The corresponding on-source observing time are 160.4, 15.6, 4.3, 3 and 6.6 minutes, and the observing bandwidths are 300~MHz, 1~GHz, 2~GHz, 2~GHz and 4~GHz. The telescope was in its B configuration. The calibrators are the same as the ones used in 2020 VLA observations, i.e., the bandpass and absolute flux calibrator is 3C 286 (J1331+305) and the phase calibrator is J1558-1409. The summary of the 2021 VLA observations is listed in Table~\ref{tab:vla_observations}.

The observations at L, S, C and X bands have been a success while P band observations were highly corrupted with radio frequency interference (RFI), especially for the first epoch of observation, thus only the result from observation taken at the second epoch (November 7) was obtained at this band. We took the same data reduction processes as what we did in 2020 observations for all bands using the Common Astronomy Software Application (CASA 5.6.2, hereafter CASA; \citealt{2007ASPC_McMullin}), with the only exception that P band observations were ionospheric corrected by applying Total Electron Content (TEC) correction; the solution files were downloaded from NASA's Crustal Dynamic Data Information System (CDDIS) webpage\footnote{\url{https://cddis.nasa.gov/Data_and_Derived_Products/GNSS/atmospheric_products.html}}. In the imaging processes, we also take the same recipe as what we did in 2020 observations, while the Briggs weighting scheme of all bands of data taken on November 7 was altered to a robust of 0.5, to compress RMS noise and make more sensitive images since much more data were flagged when compared to previous observations. 

\section{Results} \label{sec:result}
\subsection{Radio Temporal Variability and scintillation analysis} \label{subsec:variability}
Often we characterise variability of a radio light curve that is composed of a series of flux density measurements (each of flux density $F_i$ with uncertainty $\sigma_{i}$, totalling $N$ measurements) by calculating the flux variation coefficient ($V$) and the weighted reduced $\chi^2$ statistic ($\eta$) \citep[e.g.,][]{2021ApJ_Sarbadhicary}, written as: 
\begin{equation} \label{eq:V}
    V = \frac{s}{\overline{F}} = \frac{1}{\overline{F}} \sqrt{\frac{N}{N-1} \left(\overline{F^2} - \overline{F}^2\right)}
\end{equation}
and
\begin{equation}
\eta = (1/(N-1)) \; \Sigma_i^N (F_i - \xi_F)^2/\sigma_i^2
\label{eq:equation_1}
\end{equation}
In the two equations, $\overline{F}$ is the arithmetic mean of the flux density measurements $F_i$, with which the standard deviation $s$ can be worked out. $\xi_F$ is the weighted mean of $F_i$, defined as $\xi_F = {\Sigma_i^N (F_i/\sigma_i^2)}/{\Sigma_i^N (1/\sigma_i^2)}$.

The coefficient $V$ is equivalent to the fractional RMS variability or modulation index parameter ($m$), and the weighted reduced $\chi^2$ statistic ($\eta$) is the proxy of variability significance. Both quantities are used in previous transient surveys \citep[e.g.,][]{2021ApJ_Sarbadhicary, 2023MNRAS_meerkat} to diagnose steady and variable sources. We used the above formalism introduced to calculate the epoch-to-epoch long-term variability of the PRS in our VLA observations.

\subsubsection{Long-term Variability} \label{subsec:epoch_variability}
\begin{figure*}
\centering
\includegraphics[width=0.95\textwidth]{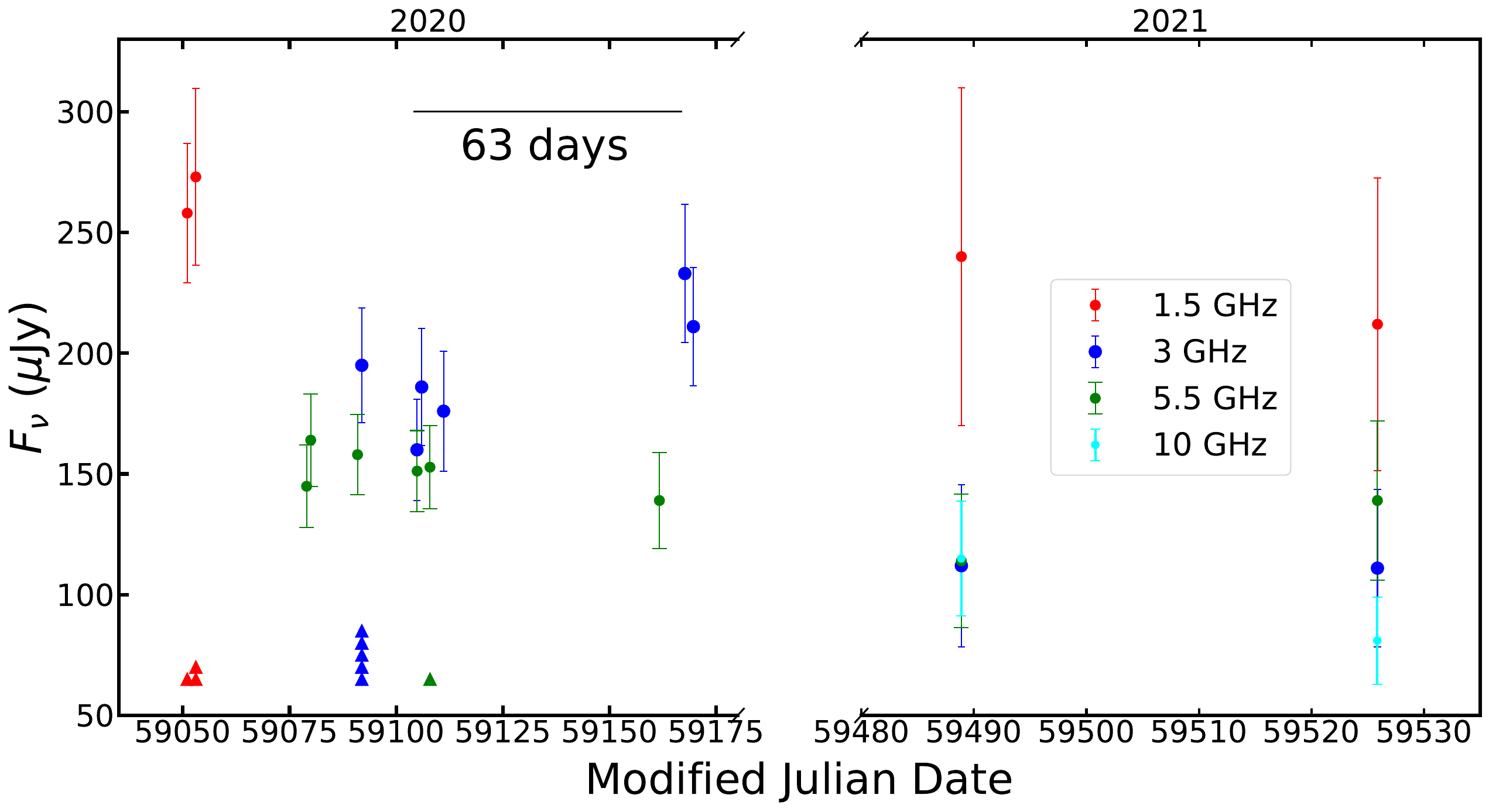}

\caption{Light curves of the PRS associated with \fastfrb~at all observing bands in 2020 and 2021 VLA campaign. %The variability is quite clear at 3~GHz on timescales of $\sim$~months to $\sim$~one year, while no significant variability is seen at other bands.
No significant variability is seen for each observing band, though the PRS at 3~GHz have varied with $\eta\sim$~2.14 in the observing campaign from 2020 to 2021. The flux of the PRS at 3~GHz rose from 160$\pm21$~$\mu$Jy on MJD 59104 to 233$\pm$29~$\mu$Jy on MJD 59167 (span of 63 days, shown by the black solid line) and decreased by a significance of the $\sim$~3.2~$\sigma$ from 193.5$\pm$10.1~$\mu$Jy to 111.5$\pm$23.7~$\mu$Jy between 2020 and 2021. The errors in the plot represent the sum of the statistical errors and 5\% additional systematic fluxscale errors. We also show in the plot the number of bursts detected at 1.5, 3 and 5.5 GHz by VLA/realfast in 2020 observations, marked with triangles at the bottom left with the same color code as that of the continuum emission. Note that the realfast system was not run on the last observation of 2020 due to a system error and was not equipped during the 2021 observations.}
\label{fig:VLA_epoch_20_21}
\end{figure*}

The epoch-to-epoch long-term source light curves of all the observing bands during the VLA campaign are shown in Fig.~\ref{fig:VLA_epoch_20_21}, in which we have added an additional 5\% systematic fluxscale error for the measured flux density to the corresponding statistical errors, which is typical for analyzing VLA observations \citep{2017ApJS..230....7P}. We have studied the variability of the PRS in the 2020 and 2021 observations and obtained that the variability significance are 0.27, 2.14 and 0.42 and the flux variation coefficients are 10.68$\pm$12.62\% and 25.33$\pm$7.05\%, 10.60$\pm$6.62\% at 1.5~GHz, 3~GHz, and 5.5~GHz, respectively. The variability significance at 10~GHz is at the level of $<$~1 over a timescale of $\sim$~one month in 2021, see Table~\ref{tab:vla_observations}. These results show that the PRS was not varied significantly at 1.5~GHz, 5.5~GHz and 10~GHz but somewhat varied marginally at 3~GHz during the VLA campaign. 

To check whether calibration problems caused the flux variation at 3~GHz, we have studied the variability of the field sources during the 2020 VLA observations and 2021 observations. We extracted sources for each epoch of observations in 2021 and from the deep image at 3~GHz in the entire 2020 observations using the package \texttt{PyBDSF} \footnote{\url{https://www.astron.nl/citt/pybdsf/index.html}} and selected sources that meet the requirements of a ``point source" introduced in ~\citep{2022Nature_Niu}. Comparing with the extracted sources in 2020, excpet for the PRS, we cross-matched in 2021 observations 13 sources in the October observation and 12 sources in the November observation, out of which only 1 source in 13 or 12 sources varied with significance of $\eta>$~3, while the PRS varied the second most significant ($\eta>$~3) among the 13 or 12 cross-matched sources. Excluding the significantly varied sources (including the PRS, both with $\eta>$~3), when fitting a straight line with slope of 1 to flux measurements of the crossed-matched sources in 2020 observations and 2021 October or November observation at 3 GHz, the obtained offsets are consistent with zero ($-10.5\pm13.6~\mu$Jy or $-8.7\pm14.0~\mu$Jy). Therefore there is no systematic change in flux between 2020 observations and 2021 October observation or November observation, this proves that there is no systematic calibration problem with the 3~GHz images produced in the 2021 observation campaign. This supports detection of apparent variability of the PRS at 3~GHz. Comparing the averaged 3~GHz flux densities in 2020 and 2021, we found a 3.2~$\sigma$ decrease from 193.5$\pm$10.1~$\mu$Jy to 111.5$\pm$23.7~$\mu$Jy. 

Besides, we also noticed that there might be a broad-band yearly flux decrease between 2020 and 2021 observations. We fixed the spectral indices to -0.4, i.e., the spectral index obtained in 2020 observations. We found the flux decrease of the normalization of the power-law model from
%from 188$\pm$5 to 147$\pm$14 $\mu$Jy at 3~GHz, a $\sim$20\% yearly and 2.8~$\sigma$~decrease in modeled flux. 
291$\pm$8 to 228$\pm$21 $\mu$Jy at 1~GHz, an overall $\sim$~20\% broad-band flux decrease. Excluding flux measurements at 3~GHz, we instead obtained a decrease of the normalization from 296$\pm$11 to 252$\pm$25 $\mu$Jy at 1~GHz -- a $\sim$~1.6~$\sigma$ yearly flux decrease, as detailed in Sec.~\ref{sec:radio_spectra}. This is additional and independent of $\sim$~3.2~$\sigma$ flux decrease at 3~GHz between 2020 and 2021. We therefore conclude that there is evidence of a $\sim$~20\% yearly flux decrease of the PRS in the VLA campaign which spans $\sim$~one year from 2020 to 2021.

Zooming in the 2020 observations, as shown in the light curve at 3~GHz (Fig.~\ref{fig:VLA_epoch_20_21} and Table~\ref{tab:vla_observations}), the PRS appeared to rise from 160$\pm21$~$\mu$Jy on MJD 59104 to 233$\pm$29~$\mu$Jy on MJD 59167, in $\sim$~63 days, which we define as a ``radio flare". Applying the fit introduced above, we found that the offset from the fit is $-6.3\pm11.3~\mu$Jy. Therefore there is no systematic change in flux between both epochs, proving the ``radio flare" variability is robust. If this is attributed to intrinsic flux increase, the upper limit on the size ($\tau$c) of the radio emitting region of the PRS variable component is $\sim$~0.14~pc based on light crossing time arguments (taking e-folding time as the light crossing time of $\sim$~170~days).

\subsubsection{Scintillation Analysis of the variability measurements}\label{subsubsec:scintillation}
It is well known \citep[][and references therein]{1990ARA&A..28..561R} that compact radio sources can show flux variability as a result of scintillation, particularly as a consequence of the small-scale inhomogeneities in the ionized component of the interstellar medium (ISM). We discuss in the following paragraph the predicted scintillation properties of the PRS as inferred with our measurements of the variability significance and flux variation coefficient above based on the formalism listed in \citep[][and references therein]{1998MNRAS.294..307W}. %Then we summarize the implications of scintillation analysis.

Using the description of \cite{1992_Narayan}, the scintillation properties have been parameterized in \cite{1998MNRAS.294..307W} as ``scattering strength": $\xi = ({\nu_0}/{\nu})^{17/10}$, $\nu_0$~is the transitional frequency which will be discussed below.
%\begin{equation}
%      \label{eqn:xi}
%      \xi = \left( \frac{\nu_0}{\nu} \right)^{17/10}
%  \end{equation}
$\xi$ = 1 corresponds to the critical value at which the ISM inhomogeneities introduce a substantial phase change, of the order of half a radian, across the first Fresnel zone, which is a characteristic property of the scintillation screen and has the angular radius of $\theta_{\rm F} = \sqrt{{c}/{2 \pi \nu D}}$ \citep[equations 2.2 and 3.4 in][]{1992_Narayan}, in which $D$~is the distance to the source.
%\begin{equation}
%  \label{eqn:theta_F}
%       \theta_{\rm F} = \sqrt{\frac{c}{2 \pi \nu D}}
%  \end{equation}
In the conventional scintillation analysis, scattering has been divided into two regimes, i.e., one is the weak regime ($\xi\ll$~1; transitional frequency $\nu_0$ $<$~$\nu$), in which there are only small phase 
changes induced by ISM over the first Fresnel zone. And the other one is the strong regime ($\xi\gg$~1, $\nu_0$ $>$~$\nu$) in which the wavefront is highly corrugated on scales smaller than the first Fresnel zone. We estimate that $\nu_0$ = 12.5~GHz ($\xi$ = 1) based on pyne2001\footnote{\url{https://pypi.org/project/pyne2001/}}, 
a python wrapper around the original FORTRAN implementation of the NE2001 Galactic free electron density model \citep{2002_Cordes1,2003_Cordes2}. So our observing frequencies (1.5~GHz, 3~GHz, 5.5~GHz, and 10~GHz) during the campaign are all belonging to a strong scattering regime. We skipped an investigation of potential scintillation at 10~GHz since the source wasn't observed at this frequency in 2020 observations and showed no variability over two epochs spanning $\sim$~one month in 2021. 

There are two main types of scintillation expected for point sources in the regime: One is the diffractive scintillation, fast and narrow-band, whose expected variation timescale is $t_{\rm d} = t_{\rm F} \xi^{-1} \sim 2 (\nu/\nu_0) ^{6/5}$~hours ($t_{\rm F}$ is the timescale for traversing the first Fresnel zone) and the width of the observing band is $\Delta \nu = \xi^{-2} \nu = \nu (\nu/\nu_0) ^{17/5}$. The expected variation timescales for \fastfrb~at all observing frequencies are within one day (ranges from $\sim$~4 hrs to $\sim$~18 hrs), while the expected bandwidths are all smaller than $\sim$~1/3~GHz (the widest, at 5.5~GHz). Thus diffractive scintillation was not expected to affect the epoch-to-epoch long-term variations of the PRS. The other is refractive scintillation, which is slow and broad-band, a case study of which is detailed in \cite{2022arXiv_Chen} for the PRS associated with FRB 20121102A. The point source variation would follow these properties in the context of refractive scintillation:

Modulation index:
\begin{equation}
      \label{eqn:mp}
      m_{\rm p} = \xi^{-1/3} = \left(\frac{\nu}{\nu_0} \right) ^{17/30}
  \end{equation}

The angular radius of the scattering disk at frequency $\nu$:
\begin{equation}
      \label{eqn:theta_r}
      \theta_{\rm r} = \theta_{\rm F} \xi =\theta_{\rm F,\nu_0} \left( \frac{\nu_0}{\nu} \right)^{11/5}
 \end{equation}

The refractive time-scale for a compact source with size smaller than the angular size of the scattering disk is:
\begin{equation}
      \label{eqn:t_r}
      t_{\rm r} \sim 2 \left(\frac{\nu_0}{\nu} \right) ^{11/5}
\end{equation}
 
%If the point source assumption fails, i.e., the source size is larger than the scattering disk, then the modulation index reduced by a factor $(\theta_{\rm r}/\theta_{\rm s})^{7/6}$, while the variability timescale increases by a factor of $\theta_{\rm r}/\theta_{\rm s}$. 
From the whole observation campaign spanning $\sim$~one year, we calculated the expected refractive scintillation parameters (the flux variation coefficient, sizes, and variability timescales) at 1.5~GHz, 3~GHz, and 5.5~GHz, as tabulated in Table~\ref{ta:predictions}. It's clear that the apparent variability of the PRS was unlikely due to scintillation as the flux variation coefficient observed at all bands are not reach the expected values. We therefore conclude that the evidence of scintillation of the PRS is lacking and the variability observed at 3~GHz should be intrinsic to the source. But if we assume scintillation is responsible for the flux variation coefficient (modulation index) and the scattering disk is 1~kpc away from us in the direction of the PRS, then we could estimate the sizes and the variability timescales of the PRS. The details are shown in Table~\ref{ta:predictions}.

From the above long-term variability study, the PRS did not show any significant variation at 1.5~GHz, 5.5~GHz, and 10~GHz ($\eta <$~1). Therefore we can not derive meaningful limits on the variability timescale as well as the source size of the PRS at these bands. But at 3~GHz, the source was somewhat varied ($\eta$~$\sim$~2.14), allowing us to obtain possible source size if it was due to scintillation. The predicted size of the scattering disk is 53~$\mu as$ at 3~GHz based on the observations separated for more than 1 year (Table~\ref{ta:predictions}). At an angular diameter distance of 809 Mpc \citep{2022Nature_Niu} for \fastfrb, 53~$\mu as$ corresponds to the size of 0.21~pc, which should be comparable to the size of the PRS to account for the observed variability at 3~GHz. But caution should be taken that the actual size could be much smaller if we assume a much more distant scattering disk in our galaxy. For example, a scattering disk at a distance of 10~kpc would mean that the size of the PRS drops down to $\sim$~0.07~pc ($\theta_{\rm r}\sim$~17~$\mu as$).

\subsection{Radio Spectral measurements and analysis} \label{sec:radio_spectra}
\subsubsection{Time-averaged and short-term radio spectra} \label{sub:radio_spectra_both}
\begin{figure*}
\centering
\includegraphics[width=0.95\textwidth]{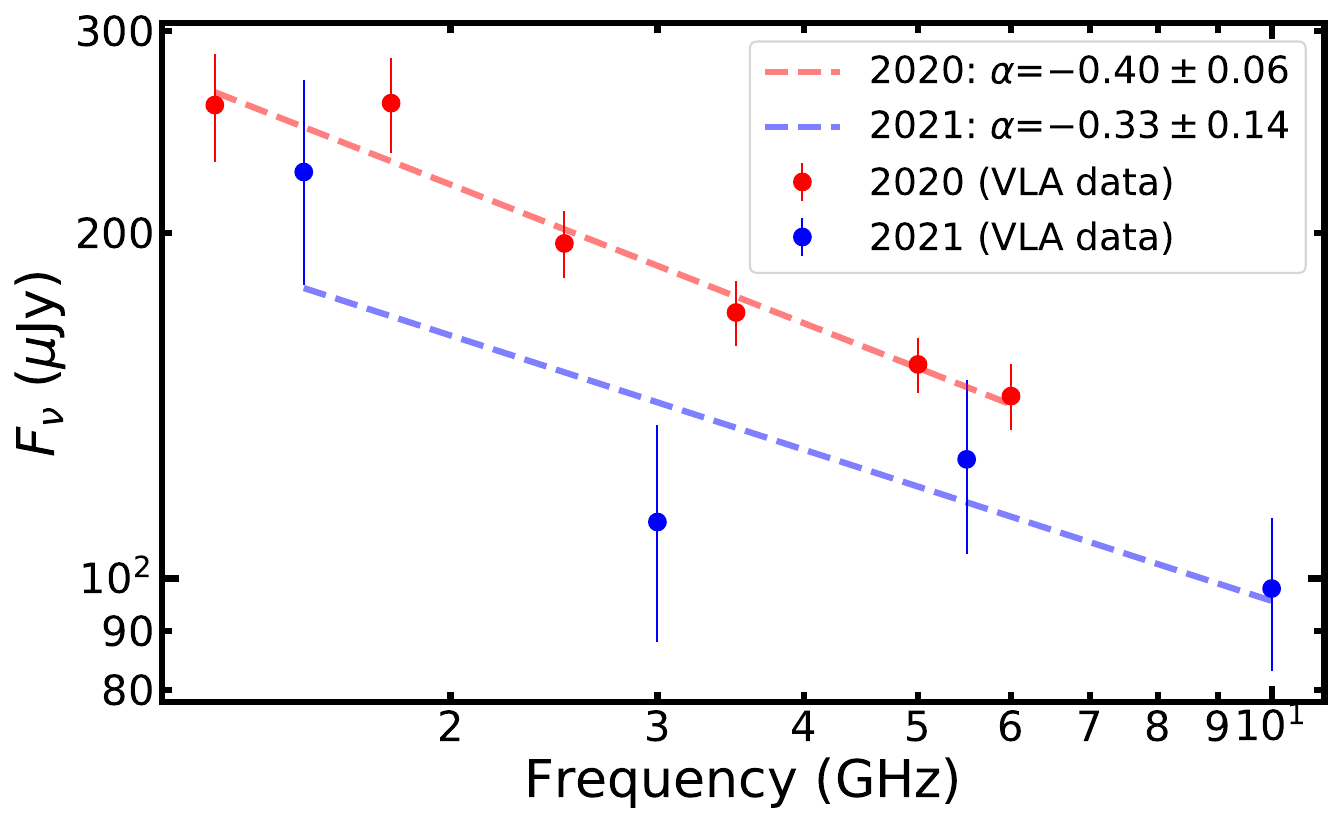}

\caption{Radio spectrum of the PRS across the 1--12~GHz frequency range in the 2020 and 2021 VLA observations. The time-averaged spectrum is well fitted by a power-law with index of -0.40$\pm$0.06 in the 2020 observations when including a 5\% VLA fluxscale systematics on measured flux densities. In the 2021 observations, the power-law index was -0.33$\pm$0.14. We observe a flux decrease between 2020 and 2021 observations. The normalizations of the two power-law spectra (indices fixed to -0.4) differ by 2.8~$\sigma$, %from 188$\pm$5 to 147$\pm$14 $\mu$Jy at 3~GHz,
from 291$\pm$8 to 228$\pm$21 $\mu$Jy at 1~GHz, suggestive of an overall $\sim$~20\% yearly flux decrease in the broadband.}
\label{fig:FRB190520B_spectra}
\end{figure*}

In the 2020 campaign, we have found the source spectrum can be well fitted by a power-law ($S_{\nu} \propto \nu^{\alpha}$, where $S_{\nu}$ is the observed flux density at the frequency $\nu$, $\alpha$ is the spectral index) with a spectral index of $-0.41\pm$0.04 \citep[]{2022Nature_Niu} averaged over $\sim$~4 months from 1~GHz to 6.5~GHz. Here we updated the fitted spectral index to $-0.40\pm$0.06 after including the statistical errors of an additional 5\% systematic fluxscale errors in the fit. In the 2021 campaign, we extended the observing band to 12~GHz as well as included P band observations at 224--480~MHz; We excluded the P band observations in our analysis in this paper since the data were largely corrupted. The 2021 observations were arranged as observing the source at each band one by one within about 3.5 hours and repeated for two epochs separated by $\sim$~1 month. The spectrum was obtained by averaging flux densities over the two epochs for each band, which yielded a power-law with a spectral index of $-0.33\pm$0.14, statistically consistent with that of the 2020 campaign (See Fig.~\ref{fig:FRB190520B_spectra} for details). We have shown that there is evidence of a yearly flux decrease of $\sim$~20\%, see Sec.~\ref{subsec:epoch_variability}, based on spectral fits with indices fixed to -0.4.  

However, the spectrum might have varied on short-term timescales, which suggests that the radio spectrum is composed of more than one component other than a single power-law which describes the observed radio spectrum. We have obtained evidence of a flat radio spectrum ($\alpha\sim$~0) in the observations on September 12, 2020 and October 1, 2021 -- the two observations out of three in the entire 2020 and 2021 observations during which multi-band observations were taken in sequence within a few hours quasi-simultaneously. Specifically, on September 12, 2020, the PRS was observed at the central frequencies of 3~GHz and 5.5~GHz, respectively. The spectrum, when we split the two bands into four sub-bands \citep[at central frequencies of 2.5~GHz, 3.5~GHz, 5~GHz, and 6~GHz;][]{2022Nature_Niu}, can be fitted with a power-law model with an index of -0.10$\pm$0.29. Remarkably, this date corresponds to the start time of the ``radio flare" we mentioned above, when the PRS appeared in a rise from 160$\pm21$~$\mu$Jy to 233$\pm$29~$\mu$Jy at 3~GHz on November 14. While the flux density at 5.5~GHz during almost the same span remained the same level (from 151$\pm17$~$\mu$Jy on September 12 to 139$\pm$20~$\mu$Jy on November 8), this indicates a spectral transition from a flat spectrum to a steep spectrum in two months between 3~GHz and 5.5~GHz. On October 1, 2021, in the upper frequency band from 2.5~GHz to 12~GHz, the flux density measurements alone reveal a spectrum with a power-law index of -0.02$\pm$0.30. Both short-term spectra above 3 GHz signatures that there might be a flat spectral component in the PRS towards higher frequencies. The spectra correspond to the two epochs in 2021 had a power-law index of $-0.23\pm$0.17 and $-0.42\pm$0.15, respectively, which does not allow us to derive independent constraints on short-term spectral variability. 

\subsubsection{Radio spectral constraints} \label{sec:radio_spec_analysis}
Self-absorption break in a spectrum of a synchrotron radio-emitting source can be used to constrain the source size and magnetic field. If the PRS radio spectrum we observed is from a single emitting component, the self-absorption break should be lower than the valid observing band at 1~GHz (L band). This allows us to put a constraint on the PRS size and the magnetic field strength.

The magnetic field strength of a synchrotron source with spectrum showing low-frequency turnover can be determined in two ways. One is through solving for the magnetic field strength dependence on source size based on the self-absorption argument at a low frequency. The other is by applying the assumption of equipartition between the energy of the radiating particles and the magnetic field. By joining them together, one can obtain the synchrotron emitting size by equating the expressions for the magnetic field strength \citep{1977MNRAS_Scott}, which writes,

\begin{equation}
\theta_{eq}=F(\alpha)(1-(1+z)^{-1/2})^{-1/17}(1+z)^{((-2\alpha+15)/34)}S_{\nu}^{8/17}\nu^{((-2\alpha-35)/34)}
\label{eq:equation_eq_theta}
\end{equation}

where $F(\alpha)$ incorporates factorial expressions $f_{1}(\alpha)$ and $f_{2}(\alpha)$, defined in \citet{1968ARA&A...6..321S}, and is a function of the spectrum index ($\alpha$) of the synchrotron source, the lowest and highest frequencies of the observations in consideration. The definition is shown in \citet{1977MNRAS_Scott}. $z$ is the redshift, $\theta_{eq}$~is the source size in arcsec and $S_{\nu}$~is measured flux (Jy) at the frequency $\nu$ (MHz). For a synchrotron source with a spectrum of the form of power-law without self-absorption turnover observed, we can make use of this equation to derive the lower limit on the source size, as well as the total energy under the assumption of equipartition of particle and magnetic field energy and thus the magnetic field. Taking the spectral index of the PRS source as -0.40 from 2020 observations and substituting the quantities into the equation above, we obtained a lower limit on the size of the PRS as $\gtrsim$~0.22~pc. %, corresponding to the brightness temperature of $T_B\lesssim$2$\times$10$^{10}$K. 
Remarkably, the value is consistent with the limit derived from ascribing the epoch-to-epoch long-term variability at S-band as due to scintillation. See Section~\ref{subsubsec:scintillation} for details. 

In addition, in the framework of equipartition assumption \citep[][and references therein]{essential2016}, the corresponding magnetic field strength for a source with integrated radio luminosity of $L$ ($\int_{\nu_{min}}^{\nu_{max}} 4 \pi D^2 F_{\nu} d\nu = 4 \pi D^2 F_{\nu_{max}} \nu_{max}^{-\alpha} (({\nu_{max}^{\alpha+1} - \nu_{min}^{\alpha+1}})/({\alpha+1}))$, which equals to $\sim3\times10^{39}$\ erg s$^{-1}$) over the total frequency range and size $R$, is,

\begin{equation}
B_{eq}=(4.5(1+\eta)c_{12}L)^{2/7}R^{-6/7}
\label{eq:equation_eq_B}
\end{equation}

where $\eta$~is the ratio between the energy of ion and electron and generally we have $\eta \sim$1 since synchrotron emission from ion is negligible, $R$ is the source size, $c_{12}$~is calculated with the spectral index and the observing highest and lowest frequencies, which correspond to 1~GHz and 12~GHz in our study (see \citet{1970ranp.book.....P} for details). In the case of the PRS, $c_{12} \approx$~1.64$\times$10$^{7}$. We therefore derive the magnetic field $B_{eq}\lesssim$~0.9~G. 
%If we attribute the dispersing region of \fastfrb~to the PRS, the value is consistent with the magnetic field strength estimated along our line of sight in \citet{2022MNRAS_Katz}, at $\sim$~500~$\mu G$, estimated by comparison of the variations of DM and Rotation Measure (RM) of \fastfrb~reported in \citet{2023Sci_Anna}.

%The total energy of the synchrotron emission of the PRS is,

%\begin{equation}
%E_{eq}(total)=c_{13}((1+\eta)L)^{4/7}R^{9/7}
%\label{eq:equation_eq_energy}
%\end{equation}

%where $c_{13}$ equals to 1.22$\times$10$^{4}$, calculated from 0.921$\times c_{12}^{4/7}$. The minimum total energy under equipartition is therefore 1.6$\times10^{47}$\ ergs.

\begin{table}
\rmfamily
\centering
\footnotesize
\begin{tabular}{|l|c|c|c|c|c|c|}
\hline
\bf Date & \bf Start Time & \bf Frequency & \bf On-Source Time & \bf Beam size & \bf Beam position angle & \bf Flux density\\
& \bf (MJD) & \bf (GHz) & \bf (min) & \bf ($^{\prime\prime}$,$^{\prime\prime}$) & (\bf $\arcdeg$) & \bf ($\mu$Jy) \\
\hline 
2020 Jul 21 & 59051.0338 & 1.5 & 96 & 4.78$\times$2.90 & -21.92 & 258$\pm$29 \\
2020 Jul 23 & 59053.0468 & 1.5 & 96 & 4.63$\times$2.90 & -16.41 & 273$\pm$37 \\
2020 Aug 17 & 59079.0050 & 5.5 & 41 & 1.38$\times$1.02 & -2.72 & 145$\pm$17 \\
2020 Aug 18 & 59079.9715 & 5.5 & 41 & 1.44$\times$1.05 & -18.38 & 164$\pm$19 \\
2020 Aug 29 & 59090.9415 & 5.5 & 41 & 1.45$\times$1.04 & -19.01 & 158$\pm$17 \\
2020 Aug 30 & 59091.9388 & 3 & 41 & 2.43$\times$1.85 & -17.99 & 195$\pm$24 \\
2020 Sep 12 & 59104.8672 & 3 & 41 & 2.76$\times$1.76 & -33.87 & 160$\pm$21 \\
2020 Sep 12 & 59104.9088 & 5.5 & 41 & 1.43$\times$1.04 & -15.89 & 151$\pm$17 \\
2020 Sep 13 & 59105.9768 & 3 & 41 & 2.44$\times$1.60 & 11.52 & 186$\pm$24 \\
2020 Sep 15 & 59107.9155 & 5.5 & 41 & 1.38$\times$1.02 & -7.94 & 153$\pm$17 \\
2020 Sep 19 & 59111.1162 & 3 & 41 & 5.00$\times$1.45 & 45.18 & 176$\pm$25 \\
2020 Nov 08 & 59161.6772 & 5.5 & 41 & 1.92$\times$0.47 & -59.53 & 139$\pm$20 \\
2020 Nov 14 & 59167.6378 & 3 & 41 & 2.76$\times$0.51 & -48.49 & 233$\pm$29 \\
2020 Nov 16 & 59169.6405 & 3 & 41 & 2.47$\times$0.48 & -46.65 & 211$\pm$25 \\ 
2021 Oct 01 & 59488.8763 & 10 & 6.6 & 0.76$\times$0.49 & -19.24 & 115$\pm$24 \\
2021 Oct 01 & 59488.8820 & 5.5 & 3 & 1.47$\times$0.92 & -19.13 & 114$\pm$28 \\
2021 Oct 01 & 59488.8851 & 3 & 4.3 & 2.41$\times$1.53 & -16.90 & 112$\pm$34 \\
2021 Oct 01 & 59488.8891 & 1.5 & 15.6 & 4.45$\times$3.05 & -15.12 & 240$\pm$70 \\
2021 Oct 01 & 59488.9005 & 0.35 & 160.4 & - & - & - \\
2021 Nov 07 & 59525.8586 & 10 & 6.6 & 0.83$\times$0.54 & 10.71 & 81$\pm$18 \\
2021 Nov 07 & 59525.8642 & 5.5 & 3 & 1.63$\times$1.03 & 11.19 & 139$\pm$33 \\
2021 Nov 07 & 59525.8673 & 3 & 4.3 & 2.69$\times$1.66 & 11.85 & 111$\pm$33 \\
2021 Nov 07 & 59525.8713 & 1.5 & 15.6 & 5.19$\times$3.41 & 13.57 & 212$\pm$61 \\
2021 Nov 07 & 59525.8828 & 0.35 & 160.4 & 20.08$\times$14.60 & 28.20 & $<$~840 \\

%2021 Oct 01 & 59488.8763 & 10 & 6.6 & 0.90$\times$0.55 & 135$\pm$18 \\
%2021 Oct 01 & 59488.8820 & 5.5 & 3 & 1.70$\times$1.03 & 177$\pm$37 \\
%2021 Oct 01 & 59488.8851 & 3 & 4.3 & 2.78$\times$1.69 & 127$\pm$20 \\
%2021 Oct 01 & 59488.8891 & 1.5 & 15.6 & 5.32$\times$3.44 & 235$\pm$49 \\

\hline 
\end{tabular} 

\caption{Summary of the VLA observations of the PRS in 2020 and 2021. The measurements of 2020 observations are taken from \citet{2022Nature_Niu}. We have included 5\% of the measured flux densities to the errors reported from CASA %reported in \citet{2022Nature_Niu} 
additionally. Frequencies refers to the central frequencies of the observing bands and beam size represents the FWHM of the corresponding synthesized beam.  The observation at P band (0.35~GHz) taken on October 1 2021 was severely influenced by RFI thus no results are included in the analysis.} 
\label{tab:vla_observations}
\end{table}

%\begin{deluxetable}{c c c c}
\begin{table}
\rmfamily
\centering
\footnotesize
\begin{tabular}{l c c c c c}
\hline

%\bf Band & \bf the flux variation coefficient & \bf Timescale & \bf Size \\
\bf$\nu$ & \bf $m_{\rm p}$ & \bf $t_{\rm r}$ & \bf $\theta_{\rm r}$ & $m_{\rm o}'$ & $\eta$ (dof) \\
(GHz) & \bf{} & (Day) & ${\mu as}$ & \bf{} & \\
\hline
1.5 & $\sim~30\%$ & $\sim 8.8$ & $\sim244$ & 10.68$\pm$12.62\% & 0.27 (3)  \\ 
3 & $\sim~45\%$ & $\sim 1.9$ & $\sim53$ & 25.33$\pm$7.05\% & 2.14 (7) \\ 
5.5 & $\sim~63\%$ & $\sim 0.5$ & $\sim14$ & 10.60$\pm$6.62\% & 0.42 (7) \\ 
\hline 
\end{tabular} 

\caption{The observed flux variation coefficient and variability significance of the PRS at 1.5~GHz, 3~GHz and 5.5~GHz in the whole VLA observations. The PRS was mostly varied at 3~GHz ($\eta \sim$~2.14), therefore we focus on the scintillation analysis at this band. Also shown here are the predicted Galactic refractive scintillation properties of the PRS assuming a scattering disk that is 1~kpc away from us, including the size of the scattering disk and variability timescale.}
\label{ta:predictions}
\end{table}

\section{Summary and Discussion} \label{sec:summary}
%\textcolor{cyan}{Summary of the properties of the PRS is ongoing.}
Based on the VLA observations of the PRS associated with \fastfrb, we measure in this paper the radio temporal variability of the source and its time-averaged and short-term radio spectra in 2020 and 2021, with inferred spectral radio luminosity, as well as constraints on the source size and the magnetic field strength assuming synchrotron self-absorption frequency below 1~GHz and equipartition of energy of particles and magnetic field. Here we summarize our results:
\begin{itemize}

      \item \textbf{Radio luminosity:} The spectral radio luminosity is ~$L_{\rm{1.5~GHz}}=4.7\times10^{29}$\ erg s$^{-1}$\ Hz$^{-1}$ and radio luminosity is $\nu L_{\rm{1.5~GHz}}=7\times10^{38}$\ erg s$^{-1}$ at a luminosity distance of 1218~Mpc in 2020 VLA observations. Similar luminosity is obtained in the 2021 VLA observations at the same observing frequency, but the modeled spectra suggest an overall decrease in flux and luminosity  of $\sim$~20\%. %See Section~\ref{sub:radio_spectra_both} for details. 
      
      \item \textbf{Temporal and spectral variability}
      We found insignificant PRS variations at 1.5~GHz, 5.5~GHz, and 10~GHz during the 2020 and 2021 campaign except that we detected marginal variability at 3 GHz in both statistical sense from epoch-to-epoch long-term measurements and direct comparison of the PRS flux. Although the average spectrum of the PRS is consistent with a single power-law with an index of -0.4, we can not exclude potential contribution of a flat spectral component since we observed spectral variation between a flat type to a steep type. The $\sim$~3.2~$\sigma$ yearly flux decreasing seen at 3~GHz combined with the spectra-modeled $\sim$~1.6~$\sigma$ flux decrease suggest a significant flux decreasing at 3~GHz. Further measurements at this band might provide critical clue to the spectral composition of the PRS. 
      
      \item \textbf{Constraints on PRS source size:} We have found no evidence of scintillation in the PRS variability studies with the VLA observations. However, if we attribute the marginal variability measured at 3~GHz as due to scintillation, then the size is obtained as $\sim$~0.21~pc when the scattering disk is set to 1 kpc. The size could be even lower by a factor of $\sim$~3 down to $\sim$~0.07~pc if the distance to the scattering disk increase by one order of magnitude to 10~kpc. If, on the other hand, there was no scintillation contribution to the variability, then the observed marginal variability at 3 GHz, including the potential ``radio flare", should be taken as intrinsic to the source. This would imply an upper limit of the size of the PRS (i.e., the variable component seen at S band) as $\sim$~0.14~pc with a light-crossing time argument. Independently, the time-averaged spectra seen in 2020 and 2021 observations, if attributed to a single spectral component, gives the lower limit of the PRS size to be 0.22~pc based on equipartition and self-absorption arguments. Our results tend to suggest that the 1--12~GHz radio emission of the PRS might consist of more than one spectral component. If so, distinct spectral components might correspond to two distinct source sizes under and above the sub-pc scale, although burst contribution to the 3~GHz PRS emission is unlikely. 
      
      \item \textbf{Magnetic field:} Radio spectral arguments give a lower limit of the PRS size of 0.22~pc as discussed above. The corresponding magnetic field strength of the PRS in the condition of equipartition is $\lesssim$0.9 G. 
      % and the total energy of source is $\gtrsim$ 1.6$\times10^{47}$\ ergs. 
      More stringent constraints would be achieved with observations taken below 1~GHz (e.g., VLA P band~$-$~0.35~GHz), which was not achieved in the 2021 VLA observations. 

\end{itemize}

The radio properties of the PRS, such as high spectral luminosity and its compact nature, as well as evidence of evolution, argue against the scenario of star-forming regions (SFR), as discussed in \citet{2022Nature_Niu}.  Pulsar Wind Nebulae (PWNe) are sources which have a typical flat spectrum ($\alpha \sim$~0) with underlying electron energy distribution of $p=1-2\alpha \lesssim$~2. In addition, such sources fade in flux density very slowly \citep[e.g., the Crab nebula,][]{1985ApJ_Aller,2007ARep_Vinyaikin}. From our spectral analysis, we found evidence of spectral variation on short time scales up to a few months. Thus it is possible that there is an underlying flat spectral component in the PRS spectrum, but then the residual spectral component in addition to a potential flat spectral component, which should be steep, should correspond to an additional origin. It is worth noting that the typical observed radio luminosity from PWNe (e.g., the Crab nebula) is several orders of magnitude lower than that of the PRS, although neutron stars with fast periods or high magnetic fields could be powering young ($\sim$~decades old) PWNe with radio luminosity orders of magnitude more luminous at early times than the PWNe observed, as predicted by models \citep[e.g.,][]{1969ApJ_Goldreich,2018ApJ_Margalit}. The PRS is also unlikely linked to Supernovae (SNe) alone either, since SNe seldom \citep[with only few cases, e.g.,][]{2021Sci_Dong} reach such a high spectral radio luminosity as the PRS, and their optically thin spectral indices are generally $<$~-0.5, steeper than the observed PRS average spectrum. Unless the PRS contains an additional flat spectral component, we can not confirm a steep radio spectral component in the PRS. It would be interesting to think about a newly-born pulsar/magnetar with on-going formation of a PWN/Magnetar Wind Nebula \citep[MWN;][]{2021ApJ_MWN} after its recent SN explosion, which might generate the observed high spectral luminosity, spectrum, and variability, as models consider rotational and magnetic energy from a young magnetar \citep[e.g.,][]{2018ApJ_Margalit,2021ApJ_MWN}. The yearly flux decrease ($\sim~$20\%), based on a constant spectral index assumption, corresponds to a power-law index of m~$\sim$~2.5 ($F_{\nu} \propto t^{-m}$) for a decade-old magnetar, which is comparable to the predicted value from model A in \citet{2018ApJ_Margalit}, in which m~$\sim$~2.2 ($m={(\alpha^2+7\alpha-2)/4}$, $\alpha=$1.3).
%($F_{\nu} \propto t^{-(\alpha^2+7\alpha-2)/4}$). 

However, the large local contribution of the DM and the decreasing trend of the DM \citep{2022Nature_Niu} as well as the swings of the RM \citep{2023Sci_Anna} indicate a dense, evolving environment of the FRB, potentially displays as the observed PRS. The dense and variable environment indicated by the locally large DM and variable DM and RM imposes crucial diagnostics to the origin of the PRS, which hardly agree with the predictions of neither PWNe nor SNe models.  It has been argued that the bursts pass through a binary wind of the FRB source (likely a magnetic neutron star) which caused dramatic sign reversal of the RM measured from bursts \citep{2023Sci_Anna}. Although source size and magnetic field are not very constraining and rather consistent, it's doubtful that such a binary wind scenario could produce the high spectral radio luminosity of the PRS. For example, \citet{2018Natur.562..233V} has shown that the radio flux of the binary wind from the Galactic X-ray Pulsar Swift J0243.6+6124 is not expected to exceed 0.01 $\mu$Jy, which equivalents to a spectral radio luminosity of $\sim 10^{21}$\ erg s$^{-1}$\ Hz$^{-1}$ at a distance of 5~kpc. Though \citet{2023A&A...673A.136R} suggest that a magnetar giant flare is capable of powering a luminous PRS ($\sim 10^{29}$\ erg s$^{-1}$\ Hz$^{-1}$) by the interaction with the stellar wind of the companion, giant flares have been seen to persist on much shorter time scales than the PRS.

The likely candidates for the PRS that are expected to generate the bright PRS radio luminosity as well as source sizes comparable to a sub-parsec scale (below and above) are relativistic jets from accreting compact objects not well-constrained in a range of mass scales. The stellar mass of the host galaxy (J160204.31$-$111718.5) of the \fastfrb~has been estimated as 6$\times$~10$^{8}M_{\odot}$ \citep{2022Nature_Niu}, which implies that it harbours an Intermediate-Mass Black Hole (IMBHs; with mass of $\sim$~10$^{2-6}M_{\odot}$) with a mass on the order of 10$^{5.99\pm0.24}M_{\odot}$ at its galactic center following the empirical relation between the mass of galaxy and the mass of the black hole \citep{2020ARA&A_Greene}. The example of the IMBH with similar radio luminosity, spectrum, and size is the one located in the dwarf galaxy SDSS J090613.77+561015.2 \citep{2020MNRAS_Yang}. Another case is the IMBH in the dwarf galaxy NGC 4395 \citep{2006ApJ_Wrobel}, though with much lower radio luminosity. But relativistic boosting effect could be mitigating the difference in the radio luminosity (i.e., 3 orders of magnitude lower). Since radio emission from such objects are generally produced by relativistically moving jets, which sometimes could be extremely relativistic (approaching the speed of light). Thus, if a jet is pointing with a small angle with respect to the line of sight, relativistic beaming effect would allow us to observe boosted jet radio emission from such systems. Depends on the type of jet, the radio luminosity would be boosted by a factor of e.g., $\delta^{3-\alpha}$ or $\delta^{2-\alpha}$, where $\alpha$~is radio spectral index and $\delta$~represents Doppler factor, defined as $\delta=\Gamma^{-1}[1-(v/c)cos\theta]^{-1}$ in which $\Gamma=(1-(v/c)^{2})^{-1/2}$ is the Lorentz factor, $c$ is the speed of light, and $v$ is the intrinsic speed of the jet making an angle of $\theta$~with the line of sight. For a relativistic jet knot (e.g., $\Gamma \sim$~5, jet intrinsic speed of $\sim$0.98~c) points to the direction with a very small angle ($\sim$~1~$\arcdeg$) with respect to the observer, the jet radio emission will be boosted by more than 3 orders of magnitude. Considering also that accreting IMBHs are able to produce yearly flux decrease or significant yearly variability, the magnetic field strength constraint ($\lesssim$~0.9~G) from our analysis is in accordance with that of IMBHs (e.g., SDSS J090613.77$+$561015.2 with magnetic field strength $\sim$~50~$\mu G$, \citealt{2023MNRAS.520.5964Y}) and the radio spectral index from an accreting compact object can produce a steep spctrum (e.g., -0.7) as well as a flat or slightly inverted spectrum (e.g., $\gtrsim$0), they can be a source of the candidate of the PRS. However, \fastfrb~is found $\sim$~1.3\arcsec away from the center of the dwarf galaxy, thus the PRS we study here, if it's associated with IMBHs, can only be associated with an off-center one with a likely mass lower than $\sim$~10$^{5.99\pm0.24}M_{\odot}$ embedded inside the dwarf galaxy, as supported by some radio observations of the dwarf galaxies \citep[e.g.,][]{2020ApJ_Reines} and some wandering massive black holes as discussed by \citet[][]{2020ApJ...895...98E,2022ApJ...927...55L,vohl2023lofar}. The PRS could be associated with a stellar mass compact object with jets under much more extreme conditions than those of the micro-quasars observed in our Galaxy \citep[][]{2004Nature_Fender,2005ApJ_Iaria,2006MNRAS_Miller-Jones}, with an apparent radio luminosity at least 10$^{5}$ times or more required for boosting, e.g., in which $\Gamma \sim$~10 or more with the inclination angle of the jet of $\sim$~0~$\arcdeg$.

\section{Conclusion}
In this paper, we investigate the temporal and spectral properties of different timescales of the PRS associated with the \fastfrb~with the VLA observations taken in 2020 and 2021. We haven found independent evidence for temporal variability of the PRS at 3~GHz and a broad-band yearly flux decrease by about 20\%. We show that the size of potential variable component in the PRS should be sub-parsec. On the other hand, based on the observed spectral properties, by assuming self-absorption and equipartition at frequencies below 1~GHz based on the lack of low frequency break of the time-averaged spectrum of the source, we derive a lower limit of the size of the PRS as a whole of $\gtrsim$~0.22~pc. We discuss some possible scenarios of the origin of the PRS. We discussed evidence neither in favor of a SFR scenario nor SNe, PWNe and binary winds origins and evidence in favor of accreting objects (e.g., IMBHs or stellar mass XRBs) are sources of the candidate of the PRS. %are much favorable candidates for the PRS based on the VLA observations as compared with other scenarios like SFRs, SNe and PWNe. 
Long-term monitoring of the PRS is promising to eventually pin down the nature of the PRS and the local environment of the FRB.

\begin{acknowledgments}
We thank the anonymous reviewer for helpful comments which has improved the manuscript significantly. We thank the staff at National Radio Astronomical Observatory for scheduling and performing the VLA observations. WY, XZ and ZY would like to acknowledge the support by the National Natural Science Foundation of China (grant number U1838203 and 11333005). 
\end{acknowledgments}


\vspace{5mm}
\facilities{VLA.}


\software{Astropy \citep{2013A&A...558A..33A,2018AJ....156..123A}, SciPy \citep{2020SciPy-NMeth}, pyne \citep{2002_Cordes1}, CASA \citep{2007CASA}, PyBDSF \citep{2015PyBDSF}.}

\clearpage




\bibliography{frb1905}{}
\bibliographystyle{aasjournal}



\end{document}